\RequirePackage{fix-cm}
\documentclass[smallextended]{svjour3}       
\smartqed  
\usepackage{graphicx}
\usepackage{mathptmx}      
\usepackage{amsmath}
\usepackage{multirow}
\usepackage{subcaption}

\begin{document}

\title{Cumulated burden of Covid-19 in Spain from a Bayesian perspective\thanks{This work was supported by grant COV20/00115 from Instituto de Salud Carlos III.}
}


\author{David Mori\~na         \and
        Amanda Fern\'andez-Fontelo \and
        Alejandra Caba\~na \and
        Argimiro Arratia \and
        Gustavo \'Avalos \and
        Pedro Puig
}


\institute{D. Mori\~na \at
              Centre de Recerca Matem\`atica (CRM) \\
              Barcelona Graduate School of Mathematics (BGSMath), Departament de Matem\`atiques, Universitat Aut\`onoma de Barcelona (UAB) \\
              Department of Econometrics, Statistics and Applied Economics, Riskcenter-IREA, Universitat de Barcelona (UB)
              \email{dmorina@mat.uab.cat}           
           \and
           A. Fern\'andez-Fontelo \at
              Chair of Statistics, School of Business and Economics, Humboldt-Universit\"at zu Berlin, Berlin, Germany
            \and
            A. Caba\~na \at
            Barcelona Graduate School of Mathematics (BGSMath), Departament de Matem\`atiques, Universitat Aut\`onoma de Barcelona (UAB) \and
            A. Arratia \at
            Department of Computer Science, Universitat Politècnica de Catalunya (UPC) \and
            G. \'Avalos \at
            Department of Computer Science, Universitat Politècnica de Catalunya (UPC) \and
            P. Puig \at
            Barcelona Graduate School of Mathematics (BGSMath), Departament de Matem\`atiques, Universitat Aut\`onoma de Barcelona (UAB)
}

\date{}
\maketitle

\begin{abstract}
The main goal of this work is to estimate the actual number of cases of Covid-19 in Spain in the period 01-31-2020 / 06-01-2020 by Autonomous Communities. Based on these estimates, this work allows us to accurately re-estimate the lethality of the disease in Spain, taking into account unreported cases. A hierarchical Bayesian model recently proposed in the literature has been adapted to model the actual number of Covid-19 cases in Spain. The results of this work show that the real load of Covid-19 in Spain in the period considered is well above the data registered by the public health system. Specifically, the model estimates show that, cumulatively until June 1st, 2020, there were 2,425,930 cases of Covid-19 in Spain with characteristics similar to those reported (95\% credibility interval: 2,148,261 – 2,813,864), from which were actually registered only 518,664. Considering the results obtained from the second wave of the Spanish seroprevalence study, which estimates 2,350,324 cases of Covid-19 produced in Spain, in the period of time considered, it can be seen that the estimates provided by the model are quite good. This work clearly shows the key importance of having good quality data to optimize decision-making in the critical context of dealing with a pandemic.
\end{abstract}

\section{Introduction}\label{intro}
SARS-CoV-2 belongs to the family of betacoronavirus and has been identified as the cause of Covid-19 disease, which can affect the lower respiratory tract and in some cases progress to pneumonia in humans. In particular, it has been identified as the causative agent of an unprecedented outbreak of pneumonia in Wuhan City, province of Hubei in China starting in December 2019\cite{ref1} and spreading rapidly all over the world and being declared as a pandemic by World Health Organization (WHO) on 2020 March 11th. Considering that many cases run without developing symptoms beyond those of MERS-CoV, SARS-CoV or pneumonia due to other causes, it is reasonable to assume that the incidence of this disease has been under-registered, especially at the beginning of the outbreak.\cite{ref2} Similarly, as many other countries’ health systems were stressed to the limit of their capacity by the pandemic, it became clear that providing researchers and general public with reliable data was almost impossible. Spain is, to the date, among the most affected European countries in terms of number of registered cases, hospitalizations and deaths and there has been a debate to what extent officially reported data can be trusted.\cite{ref3} This work aims to estimate the real burden of Covid-19 in Spain by Autonomous Community (CCAA), considering the data officially reported by the Spanish Ministry of Health (which has been reported by each CCAA health department) and to compare these estimates to the results provided by the second wave of the seroprevalence study conducted from May 18th to June 1st.\cite{ref4} In this study, 63,564 participants were recruited with a participation rate among eligible individuals around 66.5\%. Globally, the estimated prevalence of IgG antibodies against SARS-Cov-2 in Spain is around 5.2\% (95\%CI: 4.6\% - 5.4\%). 

\section{Methods}\label{methods}
A new Bayesian hierarchical framework to analyze potentially under-reported count data was recently introduced.\cite{ref5} It was originally used to estimate unreported cases of tuberculosis in Brasil, but we have adapted it to use it in the context of Covid-19 disease in Spain by CCAA. A limitation of this methodology is that the spatial effect can only be estimated on regions with at least one neighbor, so isolated CCAA cannot be included (Islas Baleares, Canarias, Ceuta and Melilla), although the incidence of the disease in these regions is much smaller than in peninsular ones. All Covid-19 cases reported by the Spanish Ministry of Health through the Instituto de Salud Carlos III by CCAA in the period 01-31-2020 to 06-01-2020 accessed by July 24th, 2020 (the data are being updated retrospectively as new information for some CCAA is available) were included in this work. The model allows for the inclusion of covariates on the true count-generating process and on the underreporting mechanism as well. Average, minimum and maximum temperature per CCAA and day (as reported by the Agencia Estatal de Meteorología\cite{ref6}) were included to evaluate their potential impact on the number of Covid-19 cases as well as an indicator for the non-pharmaceutical interventions undertaken by the Spanish government (no intervention until March 15th, declaration of the emergency state from March 16th to March 30th and from April 13th to June 1st, mandatory confinement from March 31st to April 12th). Number of PCR and antibodies tests were included as covariates that might have an impact on the underreporting mechanism. Technical details are available in Appendix A.

\section{Results}\label{results}
\begin{figure}[h!]
  \includegraphics[width=1\textwidth, height=7cm]{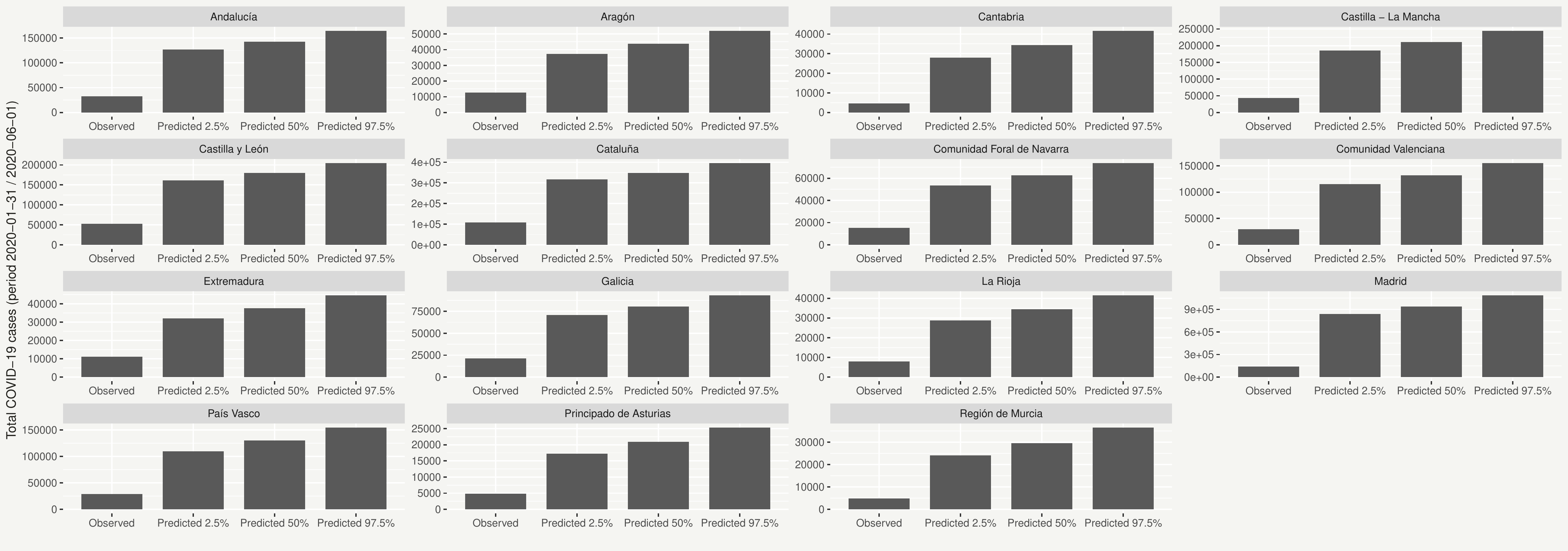}
\caption{Registered (first bar) and estimated (median and percentiles 2.5 and 97.5\% of the posterior distribution) cumulated Covid-19 cases in the period 01-31-2020/06-01-2020 in each CCAA.}
\label{fig:1}       
\end{figure}

It can be seen that the number of registered cases represent only a small fraction of the actual burden of the disease in all CCAA (Fig.~\ref{fig:1}). These unreported cases can be interpreted as asymptomatic or with mild symptoms or even cases with similar clinical characteristics than those that were registered, and the causes for un-reporting might be multiple -patients with unusual symptoms could have been misdiagnosed, limit stress of the public health system at some points of time, among others. 

By considering these unreported cases as well, it can be seen that the estimates found in this study are very similar to the results provided by the seroprevalence study conducted in Spain\cite{ref4} and that in most CCAA the projection of seroprevalence study yields 95\% confidence intervals with non-empty intersection with 95\% credible intervals (CrI) provided by the present study, as can be seen in Table~\ref{tab:1}.

\begin{table}
\caption{Registered, estimated and projected from the ENE-COVID19 study cumulated Covid-19 cases by CCAA in the period 01-31-2020/06-01-2020. CrI stands for credible interval and CI stands for confidence interval. Spain* excluding Islas Baleares, Canarias, Ceuta and Melilla.}
\label{tab:1}       
\begin{tabular}{cccc}
\hline\noalign{\smallskip}
CCAA & Registered & Estimated (95\% CrI) & Projection from  \\
     &            &                      & ENE-COVID19 \\
     &            &                      & Study (95\% CI)\\
\noalign{\smallskip}\hline\noalign{\smallskip}
Andalucía                  & 32,878  & 142,294 (126,387 – 164,403)       & 244,013 (210,356 – 286,084)       \\
Aragón                     & 12,616  & 43,877 (37,333 – 52,029)          & 64,645 (51,452 – 83,115)          \\
Cantabria                  & 4,620   & 34,282 (27,959 – 41,659)          & 18,594 (12,203 – 27,311)          \\
Castilla - La Mancha       & 43,080  & 211,286 (185,594 – 244,486)       & 209,385 (176,859 – 248,009)       \\
Castilla y León            & 52,316  & 180,408 (161,644 – 204,663)       & 179,966 (155,971 – 206,361)       \\
Cataluña                   & 108,358 & 347,729 (316,690 – 395,782)       & 468,188 (399,111 – 552,616)       \\
Comunidad Foral de Navarra & 15,326  & 62,639 (53,621 – 73,665)          & 41,870 (32,056 – 54,300)          \\
Comunidad Valenciana       & 29,302  & 132,087 (115,561 – 155,241)       & 135,102 (110,083 – 170,128)       \\
Extremadura                & 11,150  & 37,611 (31,953 – 44,561)          & 35,234 (25,625 – 46,979)          \\
Galicia                    & 21,334  & 80,723 (70,711 – 93,246)          & 59,389 (45,891 – 75,586)          \\
La Rioja                   & 7,956   & 34,389 (28,780 – 41,473)          & 12,355 (8,870 – 16,790)           \\
Madrid                     & 141,312 & 938,391 (841,130 – 1,086,685)     & 759,627 (666,339 – 866,241)       \\
País Vasco                 & 28,704  & 129,780 (109,621 – 154,185)       & 81,688 (61,818 – 108,181)         \\
Principado de Asturias     & 4,854   & 20,913 (17,238 – 25,289)          & 16,365 (11,251 – 23,524)          \\
Región de Murcia           & 4,858   & 29,522 (24,041 – 36,495)          & 23,902 (16,433 – 37,347)          \\
Spain*                     & 518,664 & 2,425,930 (2,148,261 – 2,813,864) & 2,350,324 (1,984,319 – 2,802,574)\\
\noalign{\smallskip}\hline
\end{tabular}
\end{table}
Having accurate estimates for the actual number of cases is also useful to estimate lethality associated to the disease, as it seems to be overestimated in Spain when using the officially reported cases compared to lethality estimated in other countries. Estimates for each CCAA and Spain are provided in Table~\ref{tab:2}, and it can be seen that they are much more consistent with those reported in countries with similar characteristics\cite{ref8}, all the cases being around 1\% instead of values as high as 6.84\% in Castilla La Mancha when using only registered data. The overall estimate for Covid-19 lethality in Spain is 1.10\% (95\% CrI: 0.95\% - 1.25\%).
\begin{table}
\caption{Observed and estimated lethality by CCAA and globally. CrI stands for credible interval. Spain* excluding Islas Baleares, Canarias, Ceuta and Melilla.}
\label{tab:2}       
\begin{tabular}{ccc}
\hline\noalign{\smallskip}
CCAA & Registered & Estimated lethatility (\%)  \\
     & lethality (\%) & (95\% CrI) \\
\noalign{\smallskip}\hline\noalign{\smallskip}
Andalucía                  & 4.27 & 0.99 (0.85 – 1.11) \\
Aragón                     & 6.55 & 1.88 (1.59 – 2.21) \\
Cantabria                  & 4.37 & 0.59 (0.48 – 0.72) \\
Castilla - La Mancha       & 6.84 & 1.39 (1.20 – 1.59) \\
Castilla y León            & 3.68 & 1.07 (0.94 – 1.19) \\
Cataluña                   & 5.16 & 1.61 (1.41 – 1.76) \\
Comunidad Foral de Navarra & 3.20 & 0.78 (0.67 – 0.91) \\
Comunidad Valenciana       & 4.55 & 1.01 (0.86 – 1.15) \\
Extremadura                & 4.56 & 1.35 (1.14 – 1.59) \\
Galicia                    & 2.85 & 0.75 (0.65 – 0.86) \\
La Rioja                   & 4.54 & 1.05 (0.87 – 1.25) \\
Madrid                     & 6.15 & 0.93 (0.80 – 1.03) \\
País Vasco                 & 4.96 & 1.10 (0.92 – 1.30) \\
Principado de Asturias     & 6.39 & 1.48 (1.23 – 1.80) \\
Región de Murcia           & 3.05 & 0.50 (0.41 – 0.62) \\
Spain*                     & 5.16 & 1.10 (0.95 – 1.25) \\
\noalign{\smallskip}\hline
\end{tabular}
\end{table}

\begin{figure}[h!]
  \includegraphics[width=1\textwidth, height=6cm]{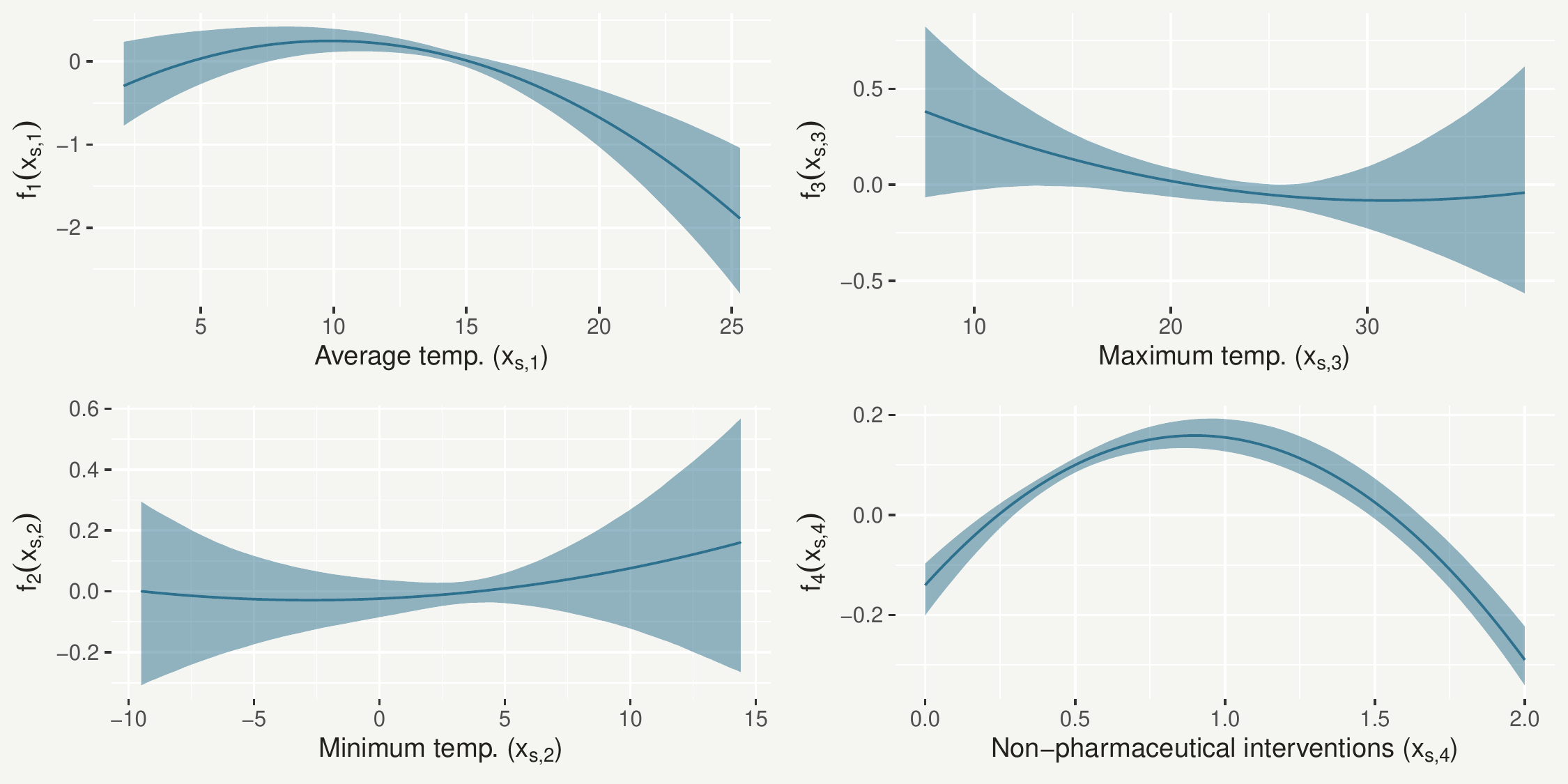}
\caption{Posterior mean predictions (solid lines) of the effects of average, maximum, minimum air temperature and non-pharmaceutical interventions on the rate of Covid-19 incidence in Spain, with associated 95\% CrIs.}
\label{fig:2}       
\end{figure}

The impact of the considered covariates on the actual Covid-19 incidence is shown in Fig.~\ref{fig:2}. It can be seen (right bottom) that the incidence rate is increasing until the declaration of the emergency state (1.0) and then decreasing drastically. Regarding the temperature effect, there is no clear pattern. Maximum and minimum temperature seem to have no effect on the Covid-19 incidence, while the decreasing incidence that can be seen when the average temperature increases disappeared in a sensitivity analysis replacing non-pharmacological interventions by a sequential indicator of time as covariate. Therefore, temperature is probably acting just as a confusing factor here.

\begin{figure}[h!]
\begin{subfigure}{0.45\textwidth}
        \includegraphics[width=0.9\textwidth]{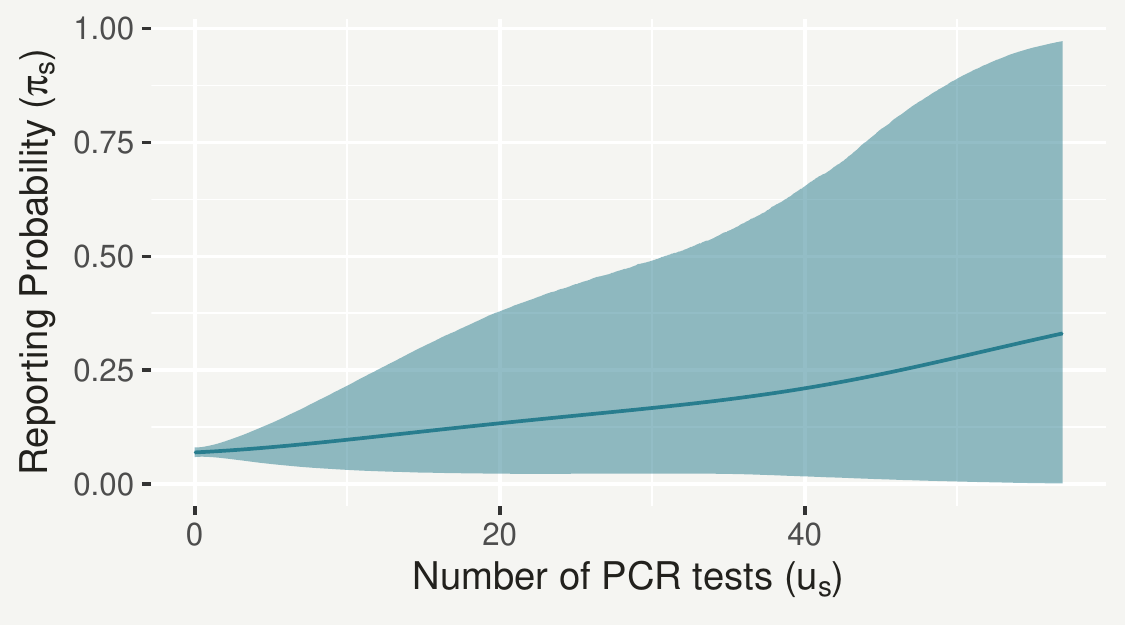} 
\end{subfigure} \hspace{0.2\textwidth}
\begin{subfigure}{0.45\textwidth}
        \includegraphics[width=0.9\textwidth]{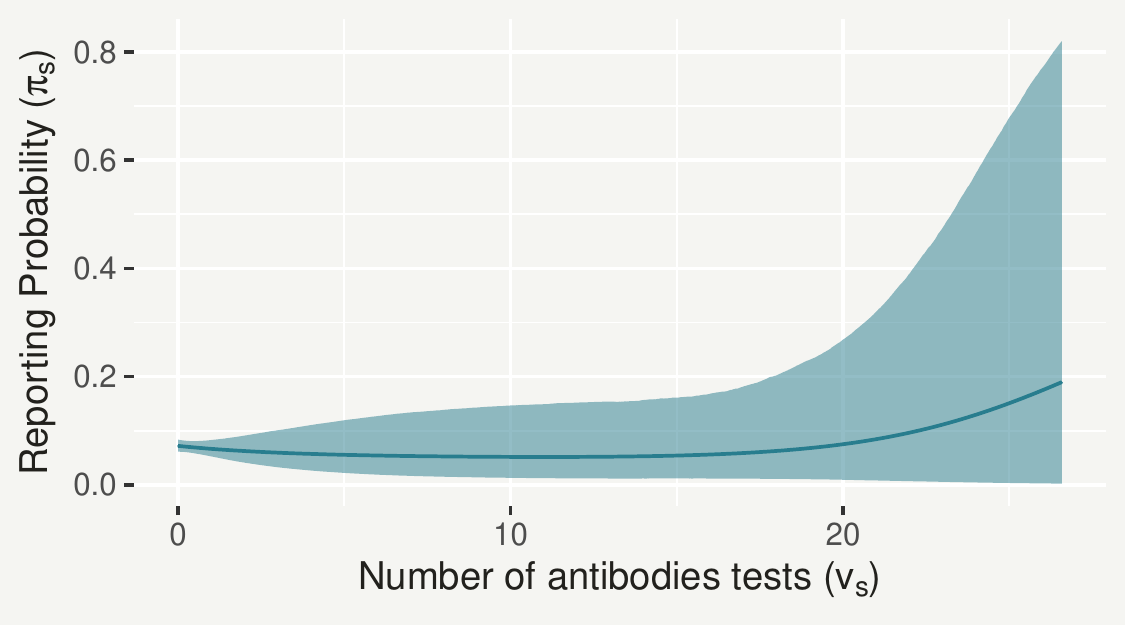} 
\end{subfigure}
\caption{Posterior mean predicted effect of PCR (left) and antibodies (right) tests on the reporting probability of Covid-19 in Spain, with associated 95\% CrI.}
\label{fig:3}
\end{figure}

Fig.~\ref{fig:3} shows that the probability of reporting a case increases as the number of performed PCR and antibodies tests increases, as could be expected.

\subsection{Model checking}\label{checking}
\begin{figure}[h!]
  \includegraphics[width=0.8\textwidth, height=6cm]{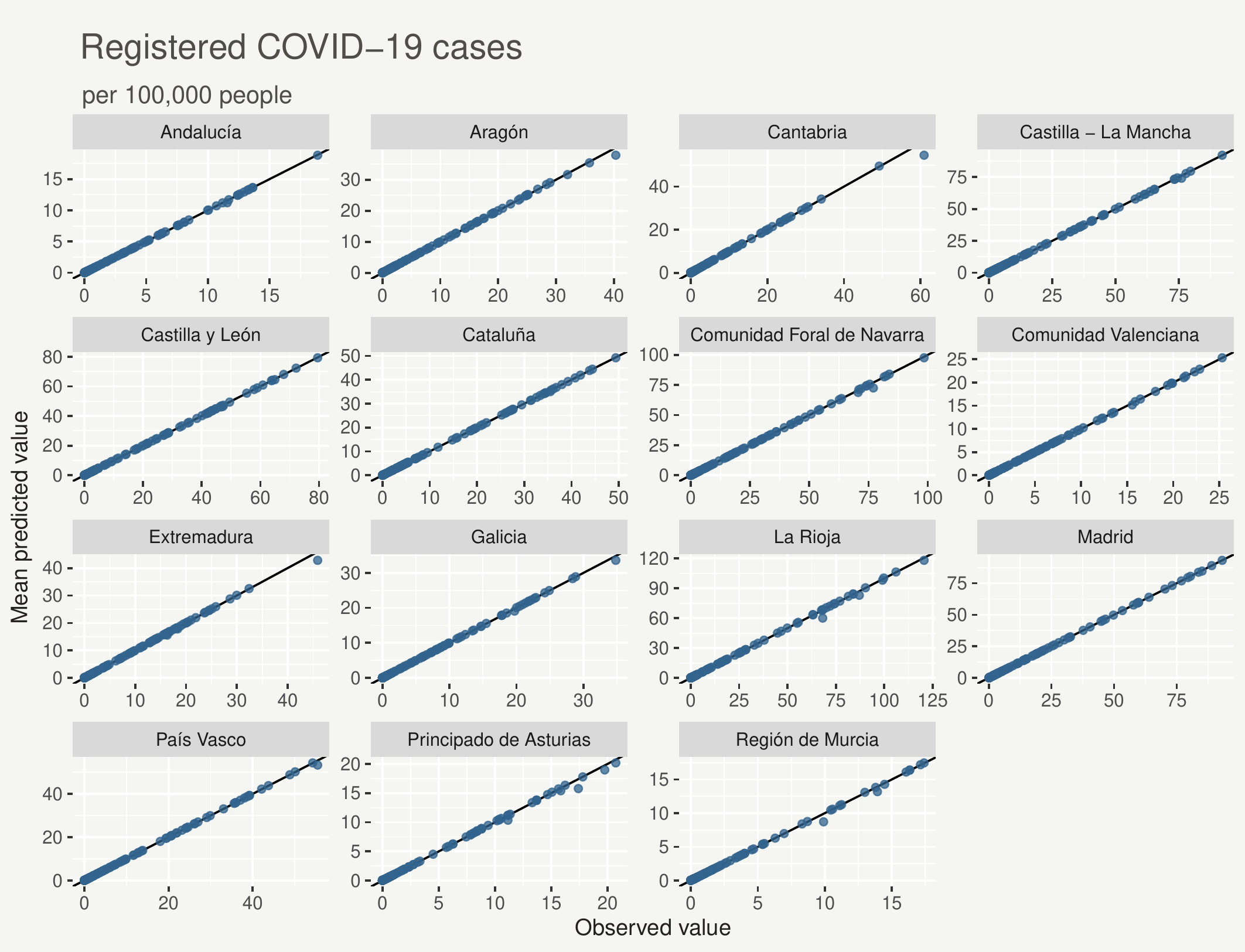}
\caption{Registered vs predicted Covid-19 cases per 100,000 individuals by CCAA.}
\label{fig:4}       
\end{figure}

The goodness of fit of the proposed model can be checked by obtaining predictions for the registered values and comparing them to the actual registered values. Fig.~\ref{fig:4} shows this comparison for each CCAA, and it can be seen that predicted and actually registered values are very similar (perfect fit would be over the diagonal). 

\begin{figure}[h!]
  \includegraphics[width=1\textwidth, height=6cm]{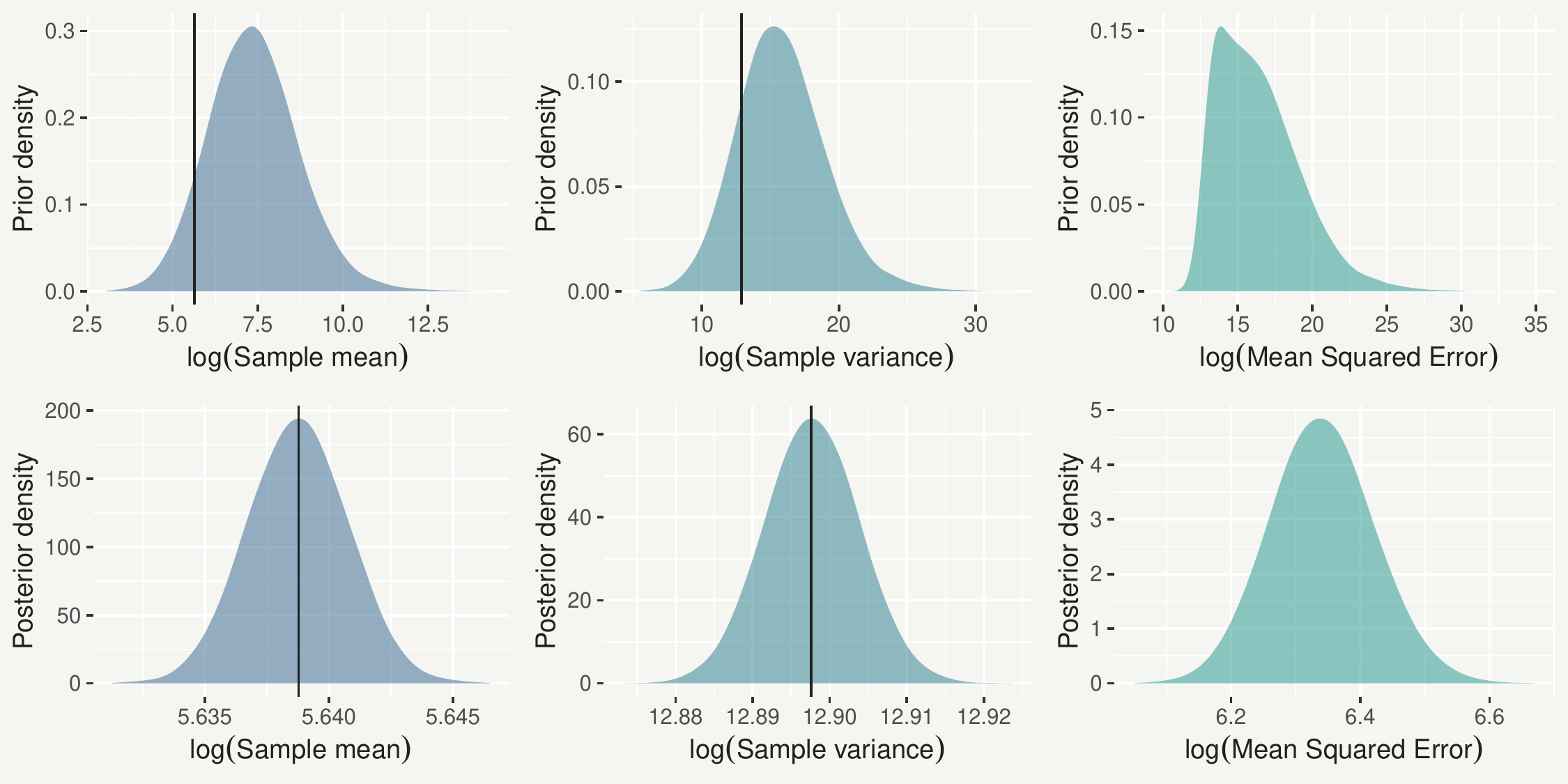}
\caption{Prior (top row) and posterior (bottom row) predictive distributions of the sample mean (left column), sample variance (central column) and the log-mean squared error from the registered counts (right column), of the replicates. Observed statistics are plotted as vertical lines.}
\label{fig:5}       
\end{figure}

The goodness of fit of the model can also be assessed by checking whether summary statistics of the registered data are fitted properly by the model through replicates. In particular, Fig.~\ref{fig:5} shows how sample mean and variance are captured by the model, comparing the prior (top) and posterior (bottom) predictive distributions of both sample statistics and the mean squared error. It can be seen that the uncertainty in the parameters has been reduced considerably by the data, as the posterior predictive distribution are notably more precise than the corresponding priors, meaning that the model is fitting the data well.

\section{Discussion}\label{discussion}
Dealing with under-reported data is very common in several fields including epidemiology, biomedical and social research among many others. It is known that predictions based on under-reported data might be severely biased if this issue is not taken into account at the modeling stage.\cite{ref9} This is especially important when dealing with diseases with a huge number of asymptomatic cases, as the Covid-19, where the proportion of infected individuals developing no symptoms can be as high as 40-45\%.\cite{ref10} This concern has received a lot of attention recently in the biomedical and methodological literature, and several proposals have been done in order to model under-reported data, from Markov chain Monte Carlo methods\cite{ref11} to time series analysis.\cite{ref9,ref12} 

This work shows that Covid-19 cases in Spain are severely under-reported and that an estimation of the unreported cases consistent with the results provided by the second wave of a nationwide seroprevalence study\cite{ref4} can be achieved by means of a hierarchical Bayesian methodology proposed very recently.\cite{ref5} The results of this study also show that non-pharmaceutical interventions undertaken by the Spanish government had a significant impact on Covid-19 incidence, as a monotonous decrease in the disease incidence following the implantation of mobility restrictions can clearly be seen. No impact of temperature could have been detected, as the apparent decrease in Covid-19 incidence for higher average temperatures was better explained by the sequential pass of time. The methodology also allows for the inclusion of covariates that might explain the under-reporting mechanism, and so, it can provide public health decision-makers with ways of improving the way data are registered. In this case, it can be seen that the more PCR and antibodies tests are conducted, the more likely is to report a case.     

It is important to notice that the considerable differences in coincidence between the estimated number of cases provided by the Bayesian methodology and the seroprevalence study can be partially explained by the differences in how CCAA reported their data to the Spanish Ministry of Health. Some of them reported prevalent cases and some included, at least lately, asymptomatic cases tested positive while others reported only cases that required some kind of medical attention, so asymptomatic and some mild symptoms cases might be missing in the data provided by these regions.  

One of the lessons that should certainly be learned from the current Covid-19 pandemic is that it is crucial to provide researchers with reliable data under extremely complex circumstances, in order to be able to assure public health decision makers are provided with the most reliable information at any time. When this is by no ways possible, the issue should be at least taken into account by using a model capable of accommodating under-reported data like the one used in this study.

\section*{Appendix A. Technical details}\label{appendix}
Let $y_{t,s}$ and $z_{t,s}$ be the actual and registered counts of Covid-19 in each Spanish CCAA $s$ and day $t$ respectively. In order to evaluate the impact of air temperature and non-pharmaceutical interventions undertaken by the Spanish government on spread of SARS-CoV-2 several covariates were considered. In particular, $x_s^1$ represents the average temperature, $x_s^2$ represents the maximum temperature, $x_s^3$ represents the minimum temperature and $x_s^4$ represents the non-pharmaceutical interventions. Additionally, the covariates $u_s$ and $v_s$ representing the number of PCR and antibodies tests undertaken respectively were considered in the characterization of the under-reporting mechanism. The variable of interest to be modelled is an indicator random variable that index the data into fully observed or under-reported. This indicator variable is to be interpreted as the proportion of true counts that have been reported, and hence is continuous in [0,1]. The model is then specified according to the following equations:
\begin{equation}
z_{t,s} \mid y_{t,s}, \gamma_{t,s} \sim Binomial(\pi_s, y_{t,s})
\end{equation}

\begin{equation}
\log\left(\frac{\pi_s}{1- \pi_s}\right)=\beta_0+g_1(u_s)+g_2(v_s)+\gamma_{t,s}
\end{equation}

\begin{equation}
y_{t,s} \mid \phi_s, \theta_s \sim Poisson(\lambda_{t,s})
\end{equation}

\begin{equation}
\log(\lambda_{t,s})=\log(P_{t,s})+a_0+f_1(x_s^1)+f_2(x_s^2)+f_3(x_s^3)+f_4(x_s^4)+\phi_s+\theta_s,
\end{equation}
where $\phi_s$ and $\theta_s$ are additive effects from a spatially unstructured random effect and a spatially structured one, respectively. Assuming that all individual occurrences have equal chance of being independently reported, $\pi_s$ can be interpreted as the probability that an actual occurrence is reported in a certain region $s$ and is in fact the indicator variable of interest. Functions $g_1, g_2, f_1, f_2, f_3, f_4$ are orthogonal polynomials of degree 2. It is known that orthogonal polynomials reduce multiple-collinearity between the monomial terms compared to raw polynomials.\cite{ref13} The term $\log(P_{t,s})$, where $P_{t,s}$ is the population, is an offset to ensure the covariates act on the incidence rate. An extra unstructured effect $\gamma_{t,s}$ was included in the model for the reporting probability (Equation (2)) to capture the effect of potential unobserved covariates on the detection rate of Covid-19.

\begin{acknowledgements}
D. Moriña acknowledges financial support from the Spanish Ministry of Economy and Competitiveness, through the Mar\'ia de Maeztu Programme for Units of Excellence in R\&D (MDM-2014-0445) and Fundaci\'on Santander Universidades. A. Arratia and G. \'Avalos acknowledge financial support from the Spanish Ministry of Science and Innovation (Contract TIN2017-89244-R) and the Centre of Cooperation for Development (CCD-UPC). This work was partially supported by grant RTI2018-096072-B-I00 from the
Spanish Ministry of Science and Innovation.
\end{acknowledgements}

%
\section*{Conflict of interest}
The authors declare that they have no conflict of interest.

\bibliographystyle{spphys}       
\bibliography{morina_fernandez_cabana_arratia_avalos_puig}   

\begin{thebibliography}{10}
\providecommand{\url}[1]{{#1}}
\providecommand{\urlprefix}{URL }
\expandafter\ifx\csname urlstyle\endcsname\relax
  \providecommand{\doi}[1]{DOI \discretionary{}{}{}#1}\else
  \providecommand{\doi}{DOI \discretionary{}{}{}\begingroup
  \urlstyle{rm}\Url}\fi

\bibitem{ref1}
C.~Sohrabi, Z.~Alsafi, N.~O'Neill, M.~Khan, A.~Kerwan, A.~Al-Jabir,
  C.~Iosifidis, R.~Agha.
\newblock {World Health Organization declares global emergency: A review of the
  2019 novel coronavirus (COVID-19)} (2020).
\newblock \doi{10.1016/j.ijsu.2020.02.034}.
\newblock \urlprefix\url{https://pubmed.ncbi.nlm.nih.gov/32112977/}

\bibitem{ref2}
S.~Zhao, S.S. Musa, Q.~Lin, J.~Ran, G.~Yang, W.~Wang, Y.~Lou, L.~Yang, D.~Gao,
  D.~He, M.H. Wang, Journal of Clinical Medicine \textbf{9}(2), 388 (2020).
\newblock \doi{10.3390/jcm9020388}.
\newblock \urlprefix\url{https://pubmed.ncbi.nlm.nih.gov/32024089/}

\bibitem{ref3}
A.~Hyafil, D.~Mori{\~{n}}a, Gaceta Sanitaria  (2020).
\newblock \doi{10.1016/j.gaceta.2020.05.003}.
\newblock
  \urlprefix\url{http://www.gacetasanitaria.org/es-analysis-impact-lockdown-on-reproduction-avance-S0213911120300984
  http://www.gacetasanitaria.org/es-analysis-impact-lockdown-on-reproduction-avance-S0213911120300984?referer=buscador}

\bibitem{ref4}
{ESTUDIO ENE-COVID19: SEGUNDA RONDA ESTUDIO NACIONAL DE
  SERO-EPIDEMIOLOG{\'{I}}A DE LA INFECCI{\'{O}}N POR SARS-COV-2 EN
  ESPA{\~{N}}A}.
\newblock Tech. rep.

\bibitem{ref5}
O.~Stoner, T.~Economou, G.~{Drummond Marques da Silva}, Journal of the American
  Statistical Association \textbf{114}(528), 1481 (2019).
\newblock \doi{10.1080/01621459.2019.1573732}.
\newblock \urlprefix\url{https://doi.org/10.1080/01621459.2019.1573732}

\bibitem{ref6}
{Agencia Estatal de Meteorolog{\'{i}}a}

\bibitem{ref8}
M.~Catal{\`{a}}, D.~Pino, M.~Marchena, P.~Palacios, T.~Urdiales, P.J. Cardona,
  S.~Alonso, D.~L{\'{o}}pez-Codina, C.~Prats, E.~Alvarez-Lacalle,
  \doi{10.1101/2020.05.01.20087023}.
\newblock \urlprefix\url{https://doi.org/10.1101/2020.05.01.20087023}

\bibitem{ref9}
A.~Fern{\'{a}}ndez-Fontelo, A.~Caba{\~{n}}a, P.~Puig, D.~Mori{\~{n}}a,
  Statistics in Medicine \textbf{35}(26), 4875 (2016).
\newblock \doi{10.1002/sim.7026}.
\newblock \urlprefix\url{http://www.ncbi.nlm.nih.gov/pubmed/27396957
  http://doi.wiley.com/10.1002/sim.7026}

\bibitem{ref10}
D.P. Oran, E.J. Topol, Annals of Internal Medicine  (2020).
\newblock \doi{10.7326/m20-3012}.
\newblock \urlprefix\url{https://www.acpjournals.org/doi/abs/10.7326/M20-3012}

\bibitem{ref11}
R.~Winkelmann, Empirical Economics \textbf{21}(4), 575 (1996).
\newblock \doi{10.1007/BF01180702}

\bibitem{ref12}
D.~Mori{\~{n}}a, A.~Fern{\'{a}}ndez-Fontelo, A.~Caba{\~{n}}a, P.~Puig,
  (2020).
\newblock \urlprefix\url{http://arxiv.org/abs/2003.09202}

\bibitem{ref13}
W.~Kennedy, J.~Gentle, \emph{{Statistical computing}} (Marcel Dekker, New York,
  1980)

\end{thebibliography}

%
%

\end{document}